\documentstyle[psfig]{caosp}
\begin{document}
\pubyear{1998}
\volume{1}
\firstpage{1}
\htitle{Effective Temperatures from Balmer Line Profiles}
\hauthor{C. van 't Veer-Menneret {\it et al.}}
\title{Effective temperatures from A5 to G5 spectral types, using Balmer line
 profiles}
\author{C. van 't Veer-Menneret,  C. Bentolila \and D. Katz }
\institute{DASGAL, Observatoire de Paris-Meudon\\France}
\date{December 29, 1997}
\maketitle
\begin{abstract}
We show how previous works (Fuhrmann et al., van 't Veer-Menneret \& 
M\'egessier) demonstrate the efficiency of the use of 
  Balmer line profiles for effective temperature determination. 
In agreement with them, we insist on the physical interest of 
this method based on the behaviour of
these lines with the variations of the parameters involved in the treatment 
of the convective transport. The comparison between Fuhrmann's results and 
ours, independently obtained, exhibits a quite good agreement. 
We show new results of effective temperature, gravity and metallicities 
for a few of our programme stars, ranging from solar to overabundant 
metallicities. 
\end{abstract}
\section{Introduction}
\label{intr}
The first step, in a detailed abundance analysis, is to obtain an
accurate effective temperature. The quality of the other fundamental 
parameters, gravity and metallicity, will follow. Reliable  
fundamental parameters are also required for internal structure models used 
for computing diffusion processes or for asteroseismology.
 
The wings of the first four members of the Balmer line series are very 
sensitive to effective temperature (hereafter $T_{\rm eff}$), and almost 
insensitive to gravity changes for $T_{\rm eff}$ less than 8500~K, 
with a limit around 5000~K, from where Stark broadening becomes 
inefficient.
Recent works (Fuhrmann et al., 1993, 1994; van~'t Veer-Menneret and
M\'egessier, 1996, hereafter VM)
have shown that Balmer Line Profiles (hereafter BLP) are also sensitive to 
the temperature structure of the models used to interpret them. This means 
that BLPs depend on the treatment of the convection transport 
and on the metallicity entering the Opacity Distribution Functions (ODF).

In section 2 we describe and comment on these previous works. In section 3 
we present our results, and compare them to Fuhrmann's ones. 

\section{Previous Works}
In a series of papers, Fuhrmann et al. (1993, 1994) and Axer et al. 
(1994) have largely demonstrated the reliability of deriving 
$T_{\rm eff}$ from the first 4 BLPs. Their most important finding can be 
summarized in this way: {\it in order to fit simultaneously the 4 observed 
profiles with the ones computed using a single model, they had to lower the 
mixing-length free parameter $\alpha$ of the convection from 1.5 down to 
0.5, its usual values being in the range 1.25, 2.0}. 
Their extensive analyses show that their finding is valid for a large range 
of temperatures and metallicities, mainly for atmospheres cooler than 6500~K, 
with solar and deficient metal content.

In fact the H$_{\alpha}$ profile was found to be insensitive to the choice of 
$\alpha$, whereas the three other lines were found to be strongly affected, 
the line depth increasing with decreasing $\alpha$.
This behaviour is due to the fact that the H$_{\alpha}$ profile is 
formed above the convective zone, and the other lines inside it, as it was 
demonstrated by Fuhrmann et al. (1993). We refer to Figures 3 and 4
of the same paper, for an extensive view of the effects on computed BLPs
of model parameters $T_{\rm eff}$, gravity, metallicity and the mixing 
length to pressure scale-height ratio $\alpha = l/H_{\rm p}$.

The main interesting consequence is that the BLPs are effective indicators 
of stellar atmosphere structure, as long as H$_{\alpha}$ and following lines 
are jointly interpreted, and constitute a direct evidence of the depth 
stratification of a stellar atmosphere.

In a more recent paper Fuhrman et al. (1997) achieved these analyses by using
the Mg Ib triplet for gravity detemination. Indeed these pressure-broadened
lines are powerful gravity indicators, assuming that $T_{\rm eff}$ has been 
determined independently and that weak Mg\,{\sc i} lines are available for Mg 
abundance determination.

In VM, with our own observational material and reduction procedures, 
we reached the same conclusions as Fuhrmann et al. (1993, 1994) about the 
influence of model structure on BLPs and consequently on derived 
$T_{\rm eff}$, as for instance for the Sun and Procyon. We have confirmed 
the necessity to lower the value of $\alpha$ down to 0.5 for a simultaneous 
fitting of H$_{\alpha}$ and H$_{\beta}$. We extend this result for hotter 
stellar atmospheres with overabundant metal content as well.
\section{Our Method and Results}
\subsection{Observations and reduction}
The observational material was described in VM.
We recall that we used the CCD receptor combined with the 
spectrograph AUR\'ELIE attached to the Coud\'e Focus of the 
152-cm reflector at the Observatoire de Haute-Provence. 
We observed in H$_{\alpha}$ and H$_{\beta}$ ranges with a spectral 
resolution of 18 000, and in the 613.5~nm and 552.0~nm ranges with a
resolution of 30 000 for abundance analyses, the signal-to-noise ratio 
being around 400 for most of the exposures.

We present in this paper some results for stars which were former targets for 
the unfortunate EVRIS space mission, and the analysis of which remains
interesting to pursue. Some of them may be candidates for future 
asteroseismology programs, others for programs concerning $\delta$ Scuti 
stars and chemically peculiar stars.

We recall (see VM) that we used Kurucz' code ATLAS9 (1993) for computing 
the models entering the calculation of profiles by the code BALMER9.
As explained in VM, the study of convection treatment in ATLAS9 showed 
that the  overshooting option used hypotheses which were not 
physically acceptable.
Then we decided to remove this option in the computation of the models we
needed. This difficulty was confirmed by Castelli et al. (1997).

We mention as an information that we have carried out, under the UNIX system,
an efficient automatisation of Kurucz' codes, allowing computations of grids 
of models, fluxes and Balmer lines.  
\subsection{Procedure}
Table~\ref{t1} summarizes the occurence of an effect on Balmer line 
broadening due to a change of the parameters involved in the model computation. 
We refer to Figure 4 in VM, which displays in what direction the variation of 
each parameter acts, and how much.
In Table~\ref{t1} we have added 3 parameters entering the mixing length 
theory of  convection: $A/V$ is the surface over volume ratio of the 
convective element, 
$y$ is a weighting factor on the optical depth in the convection 
efficiency parameter averaged on optically thin and thick cases, 
and $k$ is a factor on the convective flux taking into account the vertical 
motions of the elements. We refer to VM and Castelli et al. (1997) for more
details, comments and references about them. They are free parameters, 
which act on the BLPs in the same way as $\alpha$. 
\begin{table}[t]
\small
\begin{center}
\caption{The occurence of effects on BLPs of parameters involved in 
model computation} 
\label{t1}
\begin{tabular}{llllllllll}
\hline\hline
        & Atmosphere & & & | & Convective Zone & & & &\\
\hline
        &[M/H]   &$T_{\rm eff}$&gravity& &$\alpha$&oversh.&$A/V$&$y$&$k$ \\
\hline
H$_{\alpha}$& yes & yes & no & & no & yes & no & no & no  \\
H$_{\beta}$ & yes & yes & no & & yes & yes & yes & yes & yes \\
\hline\hline
\end{tabular}
\end{center}
\end{table}
In all cases presented here these parameters are chosen and fixed 
once and for all, only $\alpha$ was changed and the overshooting option was 
switched off. 
The metallicity and the gravity being chosen as starting values, we 
proceed as follows:
(i) first of all $T_{\rm eff}$ is derived by fitting observed and computed 
H$_{\alpha}$ profiles,
(ii) then we select the model which best fits both H$_{\alpha}$ and 
H$_{\beta}$ by adjustment of $\alpha$, (iii) the metallicity is derived by an 
iterative method through an abundance analysis which allows to determine the 
gravity and the microturbulence velocity. The results of such an iteration
appear in the first 2 rows of Table~\ref{t2}, where the 1st row gives the 
results before the iteration and the second one the results retained after 
the first and only iteration.
\subsection{Results}
\begin{table}[t]
\small
\begin{center}
\caption{Results for Procyon. The first 2 columns give the metallicity of the
ODFs and the gravities used for the starting models, $\xi_{\rm t}$ is the 
microturbulence velocity in km\,s$^{-1}$. (*) refers to Fuhrmann et al. 1997} 
\label{t2}
\begin{tabular}{llllllllllll}
\hline\hline
$[M/H]$&log $g$&$T_{\rm eff}$&$\sigma$&log $g$&$\sigma$&$\xi_{\rm t}$&$\sigma$&
  $[Fe/H]$&$\sigma$&rem. \\
  \hline
  0.00 & 4.0 & 6450K &$\pm$50 & 3.8 &$\pm$0.1 & 2.3 &$\pm$0.3& -0.1 &$\pm$0.05&
  pres. work  \\
 -0.10 & 3.8 & 6480K & & 3.9 & & 2.3 & & -0.07 & & " \\
 -0.16 & & 6470K & & 4.0 & & 2.09& & +0.01 & &  * \\
\hline\hline
\end{tabular}
\end{center}
\end{table}

\begin{table}[t]
\small
\begin{center}
\caption{Results for other programme stars. The second row gives the 
metallicity of the ODF used, $\xi_{\rm t}$ is the microturbulence velocity 
in km\,s$^{-1}$. } 
\label{t3}
\begin{tabular}{llllll}
\hline\hline
     & $\beta$ Vir & $\eta$ Cas & $\eta$ Boo & 63 Tau & $\tau$ Uma \\
  \hline
 Sp & F9V & G0IV & G0IV & Am & Am \\
$[M/H]$ & +0.1 & -0.3 & +0.3 & +0.5 & +0.5 \\
$T_{\rm eff}$ & 6000K & 5800K & 5900K & 7200K & 7050K \\
log $g$ & 4.0 & 4.2 & 3.5 & 3.8 &  \\
$\xi_{\rm t}$ & 1.5 & 1.5 & 1.5 & 3.0 &  \\
$[Fe/H]$ & +0.1 & -0.33 & +0.2 & +0.45 &  \\
\hline\hline
\end{tabular}
\end{center}
\end{table}

Table~\ref{t2} and \ref{t3} show respectively our results for Procyon, and 
for a few other stars from our programme. In all cases $\alpha = 0.5$.
Uncertainties can be summarized as follows, with [M/H] being the metallicity
in the ODFs:
$\Delta [M/H] = +0.5$ leads to $\Delta T_{\rm eff} = -200$~K, and that leads
to a decrease in the resulting [Fe/H] of -0.15 dex.
A change in $\alpha$ of 0.75 corresponds to a change in $T_{\rm eff}$ of 
about 300~K, depending on $T_{\rm eff}$ and metallicity.
\begin{figure}[t]
\centerline{
\psfig{figure=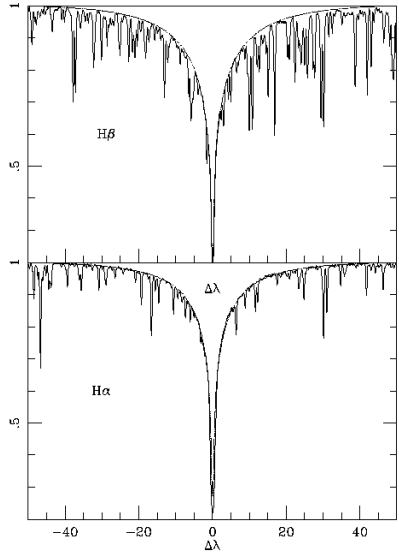,height=9.cm}}
\caption{Observed profiles (full lines) of Procyon fitted to the ones 
computed (dotted lines) with the 2nd model in Table 2}
\label{fp}
\end{figure}
Figure 1 shows, in the case of Procyon, the adequacy of the 
resulting model which represents both profiles. It is exactly as good for 
the other stars presented here. Moreover, we have to add that the ionization 
and excitation equilibria are well fulfilled, and that is the only 
disagreement with Fuhrman et al. (1997) which remains to explain. In the same 
way the discrepancy with the gravity deduced from the orbital parameters 
is not yet completely resolved.

We recall that in VM we have shown the quite good agreement between the 
$T_{\rm eff}$ derived, for the 2 Am stars, from the 2 independent methods 
BLP and IRFM. The latter gave $7175$~K for 63 Tau and $7025$~K for $\tau$ Uma,
to be compared with the values from BLPs in Table~\ref{t3}.
\section{Conclusion}
We conclude by stressing the remarkable internal consistency in the results of 
Fuhrmann and collaborators, likewise in our results, and the significant 
agreement between these two completely independant works. As a consequence, 
the value $\alpha = 0.5$  for the mixing length parameter, can be considered 
as the recommended value to be introduced in the models used for 
$T_{\rm eff}$ determination of all stars with $T_{\rm eff}$ between 8500~K 
and 5000~K, when H$_{\alpha}$ is not available.

\end{document}